\PassOptionsToPackage{prologue,dvipsnames}{xcolor}
\documentclass[UKenglish]{article}
\let\SUP\textsuperscript
\usepackage[T1]{fontenc}
\usepackage[UKenglish]{babel}
\usepackage{graphicx} 
\usepackage{amsmath}
\usepackage{amssymb}
\usepackage{relsize}
\usepackage{tikz-qtree}
\usepackage{tikz}
\usetikzlibrary{trees,arrows}
\usepackage{verbatim}
\usepackage{braket}
\usepackage{xargs}
\usepackage{xcolor}
\usepackage{todonotes}
\usepackage{xifthen} 

\usepackage{etoolbox} 
\usepackage[title]{appendix}
\usepackage{afterpage}
\usepackage{flafter}
\usepackage{float}
\usetikzlibrary{calc, shapes, backgrounds}
\usetikzlibrary{shapes.geometric, arrows}
\usetikzlibrary{calc, shapes, backgrounds}
\usetikzlibrary{shapes.geometric, arrows}
\usetikzlibrary{automata, arrows.meta, positioning}
\usepackage{tabu}

\usepackage[utf8]{inputenc}
\usepackage[acronym]{glossaries}
\usepackage{csquotes}
\usepackage{soul}
\usepackage{xparse}    
\usepackage{hyperref}  
\usepackage{nameref}   

\newcolumntype{L}{>{\centering\arraybackslash}m{5cm}}
\newcolumntype{Z}{>{\centering\arraybackslash}m{1.6cm}}
\usepackage{todonotes}
\usepackage{xspace}

\NewDocumentCommand{\sectionref}{m o}{%
    \IfValueTF{#2}
      {subsection~``\nameref{#1}''~in~\nameref{#2}\xspace}
      {``\nameref{#1}''\xspace}
}
\usepackage[textwidth=17cm,textheight=21.7cm]{geometry}
\usepackage{tikz}
\usetikzlibrary{quantikz2}
\usepackage[shortlabels]{enumitem}
\usepackage{diagbox}
\usepackage{graphicx}
\usepackage{subcaption}
\usepackage{booktabs}
\usepackage{xr}
\usepackage[backend=biber,style=nature]{biblatex}
\title{Comparing One- and Two-way Quantum Repeater Architectures}

\author{Prateek Mantri\SUP{*}, Kenneth Goodenough, and Don Towsley\\ \small
\SUP{*}\textit{Corresponding Author}: \href{mailto:pmantri@cs.umass.edu}{pmantri@cs.umass.edu}}
\date{\textit{Robert and Donna Manning College of Information and Computer Sciences,}\\\textit{University of Massachusetts, Amherst, MA, USA 01002}}

\begin{document}

\maketitle

\begin{abstract}
Quantum repeaters are an essential building block for realizing long-distance quantum communications. However, due to the fragile nature of quantum information, these repeaters suffer from loss and operational errors. Prior works have classified repeaters into three broad categories based on their use of probabilistic or near-deterministic methods to mitigate these errors. Besides differences in classical communication times, these approaches also vary in technological complexity, with near-deterministic methods requiring more advanced hardware. Recent increases in memory availability and advances in multiplexed entanglement generation motivate a fresh comparison of one-way and two-way repeater architectures.

In this work, we present a two-way repeater protocol that combines multiplexing with application-aware distillation, designed for a setting where sufficient high-quality memory resources are available --- reflecting architectural assumptions expected in large-scale network deployments. We introduce a recursive formulation to track the full probability distribution of Bell pairs in multiplexed two-way repeater architectures, enabling the performance analysis of multiplexed repeater schemes which use probabilistic $n$-to-$k$ distillation. Using this framework, we compare the proposed two-way protocol with one-way schemes in parameter regimes previously believed to favour the latter, and find that the two-way architecture consistently outperforms one-way protocols while requiring lower technological and resource overheads.
\end{abstract}


\section{Introduction}

Quantum communication is poised to enable transformative applications in quantum sensing~\cite{degen_quantum_2017, eldredge_optimal_2018, ge_distributed_2018, guo_distributed_2020}, distributed quantum computing~\cite{caleffi2024distributed, de_andrade_universal_2023}, secure communications~\cite{pirandola_advances_2020, pan2024evolution}, and quantum secret sharing~\cite{hillery_quantum_1999, markham_graph_2008}, among others. As quantum processors grow in sophistication, a scalable network for quantum information transfer between spatially separated nodes is becoming increasingly critical.
 
However, transmitting quantum information over long distances poses significant challenges, primarily due to losses that grow exponentially with distance in optical fibers. Unlike classical communication, quantum networks cannot employ classical `receive and re-transmit' strategies because of the no-cloning theorem, which prohibits the duplication of unknown quantum states. Prior studies have established fundamental limits on direct quantum information transmission~\cite{pirandola_fundamental_2017, takeoka_fundamental_2014}, underscoring the need for innovative solutions such as quantum repeaters.

Quantum repeaters are specialised devices designed to extend the range of quantum communications by dividing long segments into shorter segments, thereby mitigating losses through specific quantum gate and measurement operations. These devices significantly enhance the viability of long-distance quantum communications by employing shorter, manageable links to create extended connections. However, the implementation of these operations is fraught with errors, which can restrict the effective distance for practical quantum communication. To address both loss and operational errors, researchers have proposed various probabilistic (heralded generation and distillation operations) and near-deterministic (quantum error correction) approaches. These strategies have led to the development of three generations of quantum repeaters, each with distinct characteristics and technological requirements~\cite{munro_inside_2015, muralidharan_optimal_2016}. 

First generation architectures use probabilistic entanglement generation to mitigate loss errors, and heralded entanglement purification (often probabilistic) to mitigate operational errors. Second generation architectures also use probabilistic entanglement generation but incorporate near-deterministic quantum error correction to address operational errors. Third generation architectures solely rely on near-deterministic quantum error correction schemes to correct both operation and loss errors. The probabilistic solutions, while more feasible with current technology, necessitate heralding and consequently suffer from increased temporal costs associated with classical communication. In contrast, quantum error correction-based approaches typically employ one-way signalling and have the potential to achieve higher secret-key rates. In this manuscript, we focus on first generation (referred to in this manuscript as the two-way schemes) and third generation (referred to as the one-way schemes) quantum repeater architectures. As a special case, we also include a particular variant of second generation networks that aims to use multiplexing to establish at most a single elementary link between the segments. However, this approach does not employ error correction to protect against operation noise.

To be viable, one-way schemes typically require almost perfect operations, along with complex encoding and decoding circuitry with a large number of measurement and gate operations. Moreover, to be able to correct for fibre loss errors, one-way schemes also require repeaters to be more closely spaced compared to two-way schemes~\cite{muralidharan_ultrafast_2014, niu_all-photonic_2023}. This trade-off between complexity and performance motivates a re-examination of both approaches across relevant parameter regimes. In particular, it is important to assess whether the purported gains of one-way schemes --- despite their high technological and resource demands --- can instead be realised using the comparatively simpler two-way architectures.

This motivates a thorough comparison between different quantum repeater approaches. In their pivotal work, Muralidharan \textit{et al.}~\cite{muralidharan_optimal_2016} performed a foundational comparison of the three generations of quantum repeaters, identifying coupling efficiency ($\eta_c$), gate errors ($\epsilon_G$), and gate times ($t_G$) as critical parameters for evaluation. They delineated the specific parameter regimes in which each generation is expected to excel. However, as quantum technologies advance, revisiting these comparisons with updated models and technologies is essential to ensure accurate assessments and practical guidance for implementation.

One such example of advances in quantum technology has been in the area of long-lived memories~\cite{noauthor_next_2020, wang_single_2021, dudin_light_2013} --- a crucial requirement for the viability of two-way schemes over long distances. Unlike one-way repeaters, which may only need such memories in the case of slow gate operations, two-way protocols necessarily depend on memories that outlive the round-trip classical communication time for heralded feedback. Most prior analyses of two-way repeater schemes have focused on scenarios where memory availability is highly constrained, making memories the most significant cost factor~\cite{razavi_quantum_2009, muralidharan_optimal_2016}. This focus has shaped the strategy of using multiplexing to maximise the success probability of at least one successful link per segment~\cite{guha_rate-loss_2015, muralidharan_optimal_2016, dhara_subexponential_2021, dhara_multiplexed_2022}. This kind of scheme has also been referred to as the second generation without encoding or ‘2G-NC’~\cite{muralidharan_optimal_2016}.

Some studies have explored more aggressive forms of multiplexing that aim for multiple simultaneous successes per segment~\cite{collins_multiplexed_2007, razavi_physical_2009, razavi_quantum_2009, huie2021multiplexed, rozpedek_all-photonic_2023, li2025generalized}. However, most of these studies analyze throughput in the context of linear-chain quantum networks with probabilistic swapping realised through optical circuits, but without any distillation capabilities. In such networks, a nested swapping schedule provides limited benefit. In the cases where nested distillation schemes have been considered in the context of multiple-success multiplexing, deterministic distillation protocols are often used to simplify calculations~\cite{razavi_quantum_2009, razavi_physical_2009}. Some protocols also employ blind operations --- where the repeater proceeds without waiting for classical information, typically at the cost of an exponentially decreasing success probability --- when considering a nested swapping schedule, simplifying the computation of the number of Bell pairs delivered and the time required~\cite{razavi_physical_2009, razavi_quantum_2009}. To the best of our knowledge, no prior work has performed a detailed comparison between one-way and multiplexed two-way protocols under such considerations.

In this manuscript, we investigate the performance of multiplexed two-way (first-generation) and one-way (third-generation) quantum repeater architectures in parameter regimes where one-way schemes have previously been considered advantageous. We consider a nested swapping protocol tailored for a setting in which a sufficient number of high-quality quantum memories are available --- reflecting the architectural assumptions expected in practical large-scale network deployments. This protocol supports flexible distillation scheduling optimised for different service metrics, such as secret-key rate or fidelity thresholds. We aim to provide an even-handed performance comparison between these architectures under conditions that reflect large-scale network implementations. To support this, we introduce a recursive numerical framework that captures the full probability distribution of the number of available Bell pairs at each stage of the protocol. This framework enables a fine-grained comparison of architectures using performance metrics relevant to real-world deployment.

\section{Methods}\label{sec:Methods}

\subsection{Two-way repeater architecture}

\begin{figure}[!htb]
    \centering
    \frame{\includegraphics[width=\textwidth]{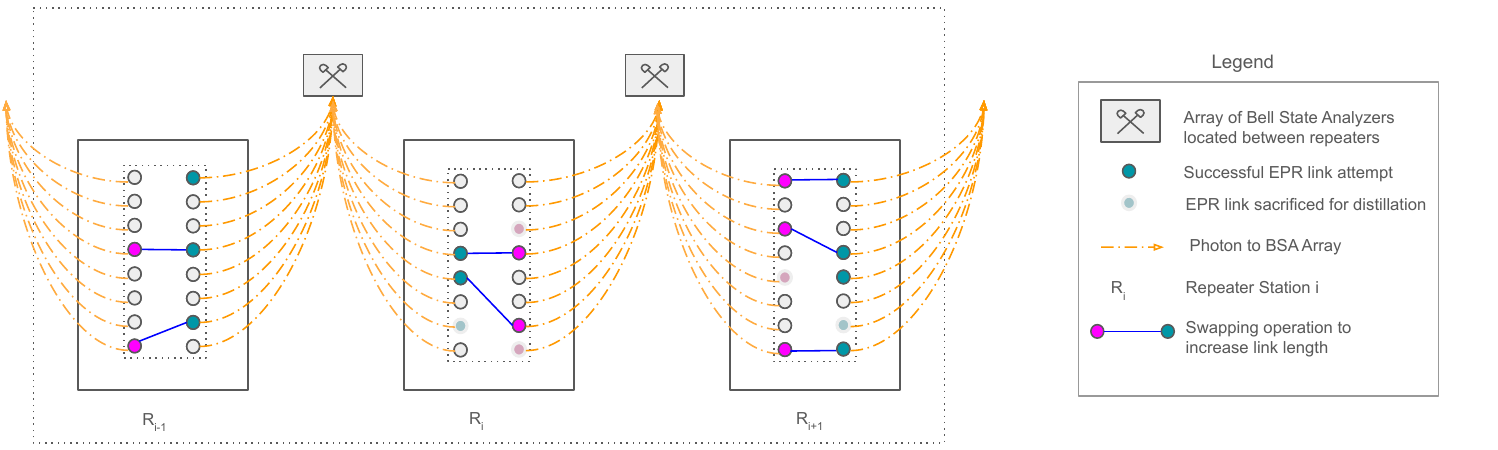}}
    \caption{\textbf{A schematic of the multiplexed Two-way repeater scheme}. Each repeater (denoted $R_i$ etc.) has multiple emitter memories located on both sides. Each of these memories emit photons creating a spatially or spectrally multiplexed burst. At the mid-point between any two repeaters, an array of Bell state analyzers exist to entangle photons coming from both sides. Only a fraction of photons from either side survives the journey and reach the mid-point station.}
    \label{fig:sys_desc}
\end{figure}

In this manuscript, we consider a linear network with each repeater station being equipped with a large number of optically active memories or emitters. These memories emit photons, which are then sent to a station located at the midpoint of the link connecting the two repeater stations. At this midpoint station, photons from two different repeaters are entangled and measured together, effectively creating a Bell pair link shared between the repeaters (See Figure \ref{fig:sys_desc}).

\subsubsection{Multiplexing for exactly one success across the network (2G-NC)}
Multiplexing is a well known technique in telecommunications and computer networks, where multiple information channels are combined over a shared medium. Multiplexed quantum repeaters can be used to overcome the probabilistic loss errors associated with signal decay in optical fibers. This is achieved by attempting multiple Bell pair generation attempts in parallel through either spatial, time-bin, or frequency multiplexing. Due to poor rates associated with entanglement generation sources, various proposals have been made over time for the use of multiplexing to improve rates in quantum repeaters ~\cite{munro_quantum_2010, piparo_resource_2020, chen_zero-added-loss_2023}. The most basic of these proposals involve parallelizing operations and sending multiple photons over an optical fiber (using time-division, spectral or spatial multiplexing), performing Bell State Measurements (BSMs), effectively creating entanglement between multiple matter qubits. To simplify analysis, these techniques often focus on using multiplexing to maximise the success of at least one Bell pair, with only one Bell pair kept between repeaters in the event of multiple successes. Moreover, this technique requires spatial or temporal multiplexing, which are often realised using lossy optical switches adding further loss~\cite{dhara_subexponential_2021}. However, a recent proposal by Chen \textit{et al.}~\cite{chen_zero-added-loss_2023} uses spectral multiplexing and parallel entanglement creation to achieve high transmission rates without the added losses usually incurred from spatial or temporal multiplexing methods. 

Another line of thought has explored multiplexing for quantum networks in multiple degrees of freedom (DOF) of a single photon~\cite{piparo_quantum_2019, piparo_resource_2020}. This scheme uses a single photonic pulse to entangle multiple pairs of remote memories, minimizing the need for extensive spatial channels and precise temporal coordination. These proposals simplify the infrastructure needs while enhancing the rate at which entangled pairs can be generated in a quantum network. Furthermore, these techniques have also been extended to one-way schemes~\cite{nishio_resource_2023, nishio2025multiplexed}.

In Muralidharan \textit{et al.} 2016~\cite{muralidharan_optimal_2016}, the authors categorise the use of multiplexing in a two-way protocol as a `second generation without encoding’ (2G-NC) scheme. This approach primarily aims to improve the probability of generating exactly one Bell pair between neighbouring repeater stations (see Equation~\eqref{eq:2gnc_prob_generation} in subsection~``\nameref{subsec:subsec_2gnc}''). It has been shown to be effective in regimes with moderate to low gate errors and low coupling efficiency. To benchmark the multiplexed two-way protocol considered in the manuscript, we have conducted a comparative analysis using the 2G-NC protocol (see~\nameref{sec:Results}). Consistent with the formulation in~\cite{muralidharan_optimal_2016}, we do not incorporate entanglement distillation into the 2G-NC scheme (see subsection~``\nameref{subsec:Distillation}'').


\subsubsection{Multiplexing for More Than One Success}\label{subsec:multiplexing}

In this manuscript, we primarily focus on multiplexing schemes that allow for the creation of multiple Bell pairs across segments. The emphasis lies in using multiplexed channels to generate multiple elementary links and not just to boost the success probability of single elementary link generation. This effectively reduces the inefficiencies associated with the basic multiplexing scheme. In our setup, we consider generating these multiplexed links across segments at the same time or with insignificant time delay. By insignificant time delay, we mean that the time difference between entangled photons arriving at the midpoint station is significantly smaller than the elementary link propagation delay and memory decoherence times, as may be achieved through spatial, spectral, or time-division multiplexing. This corresponds to an idealised synchronisation of photon arrival times at the midpoint station. While this assumption simplifies the analysis and is common in theoretical models~\cite{duan_long-distance_2001, sangouard_quantum_2011, muralidharan_optimal_2016}, achieving such synchronisation experimentally remains challenging. However, recent progress in photonic memory and emission timing control~\cite{davidson2023single, zhang2024realization} indicates that synchronised operation is a realistic target for near-future implementations.

\begin{figure}[!htb]
    \centering
    \includegraphics[width = \textwidth]{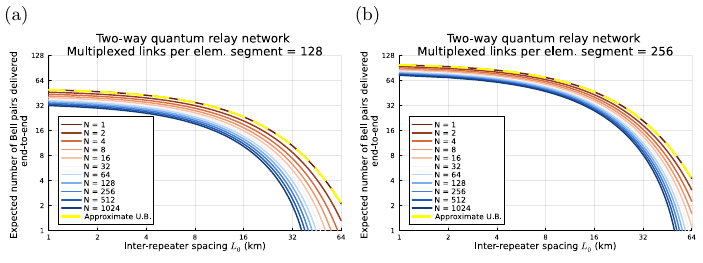}
    \caption{\textbf{Expected number of surviving Bell pairs per shot versus inter-repeater distance in a Quantum relay network.} Panel (a) shows results for 128 multiplexed elementary links; panel (b) shows results for 256 multiplexed elementary links. The y-axis denotes the expectation of the minimum number of Bell pairs generated across all segments. With deterministic swapping operations, the minimum number of Bell pairs across segments also translates into the number of Bell pairs delivered end-to-end. The x-axis denotes the inter-repeater spacing, and $N$ denotes the number of segments. The yellow dashed line shows an upper bound used by prior analyses to approximate the number of Bell pairs delivered.}
    \label{fig:expectation_mux}
\end{figure}

This enhancement allows us to balance the qubit resources required in one-way repeaters vis-\`a-vis two-way repeaters and to more accurately estimate end-to-end performance by tracking the full probability distribution of successful links. While multiplexing can enhance delivery rates, the number of end-to-end Bell pairs that can be delivered decreases as the number of segments increase in a linear relay network. Figure~\ref{fig:expectation_mux} shows the expected number of end-to-end Bell pairs that can be delivered in a single shot for a linear quantum relay network with deterministic swapping operations, for varying number of segments. The yellow dashed line denotes an approximate number of end-to-end Bell pairs by the quantity $M\cdot \pi_0$, where $M$ is the multiplexed channels and $\pi_0$ is the elementary link success probability. This quantity has been used by some prior analyses~\cite{razavi_physical_2009} as an upper bound for the expected Bell pairs a quantum relay network can deliver. As evident from Figure \ref{fig:expectation_mux}, keeping track of the probability distribution allows us to provide a more precise expectation of output Bell pairs than the models considered in prior works.

Another important consideration is that end-to-end links created using a relay approach will potentially suffer a decay in fidelity owing to swapping operations --- we address this in subsection~``\nameref{subsec:LinkPropagation}''). To deliver as many high-fidelity end-to-end Bell pairs as possible, one may need to consider distillation operations (see subsection~``\nameref{subsec:Distillation}''). However, distillation schemes like DEJMPS are inherently probabilistic with success rates dependent upon the fidelity of the input Bell pairs. This makes an exact analysis difficult. To calculate and optimise repeater schemes, it is important to determine the probability distribution of the number of successfully distilled pairs at each step. We model the number of available Bell pairs as a random variable, with a distribution affected by non-deterministic operations (See equations \eqref{eq:basic_prob_generation} and \eqref{eq:p_prime_i} in subsection~``\nameref{subsec:recursive_calcs}'').  

In subsection~``\nameref{subsec:recursive_calcs}'', we present a recursive formulation that builds upon the idea of tracking probability distributions and incorporates additional elements such as distillation (see subsection~``\nameref{subsec:Distillation}'') and nested swapping (see subsection~``\nameref{subsec:LinkPropagation}''). This formulation provides a more comprehensive framework for modeling the dynamics of Bell pair generation, distillation, and swapping, allowing for a detailed analysis of the overall system performance. By integrating these elements, our recursive approach offers a significant improvement over previous models, enabling more accurate predictions and better optimization of quantum communication protocols.

\subsubsection{Elementary Link Generation}\label{subsec:elementary_link_generation}
In this setup, we use a meet-in-the-middle protocol with spatial (or spectral) multiplexing. Each repeater has an ensemble of emitters located on either side. This ensemble has a generation frequency $\nu$, where it generates $M$ photons entangled with $M$ emitters in $\nu^{-1}$ time. This generation cycle of producing $M$ multiplexed pairs in $\nu^{-1}$ time is referred to as a \textit{burst}. One photon from each entangled photon pair thus generated is then sent to a Bell state analyser located exactly midway between any two repeater stations, hereinafter referred to as the midpoint analyser. 
We further assume emissions are synchronised across the chain, and the photons arrive at the midpoint station at the same time, and are able to effectively remove the `which-path-information'. The midpoint analyser then performs parallel BSMs on all $M$ incoming photons from each side with a success probability of $1/2$, and communicates the result to both repeater stations. There is also a need for an optical switch to separate the various multiplexing modes to the respective detectors at the analysers. This switching operation could be lossy depending on the multiplexing or the detectors used. Recent advances in detection technology~\cite{miki_64-pixel_2014} have shown promise for building large detector arrays with spatial resolution that can potentially allow us to forego the need for optical switches.
While an explicit analysis has not been done, our proposed protocol is also compatible with a mid-point source (MPS) scheme like Zero-Added Loss Multiplexing (ZALM)~\cite{chen_zero-added-loss_2023}, and will provide similar results. We also assume that time-bin dual-rail encoding is used for each multiplexed channel primarily since it allows for protection against depolarization of the photon in the channel. However, if a polarization-based encoding is used, the elementary link generation equations will have to be updated to accommodate relevant noise models.


\subsubsection{Link Propagation}\label{subsec:LinkPropagation}

Two-way repeaters use a swap operation for link propagation. A swap operation involves performing a controlled-NOT (CNOT) gate on the halves of two Bell pairs situated at a middle repeater, and measuring the involved qubits at the middle repeater to create a longer link. Depending upon the technology used, this swap operation may be probabilistic or deterministic, however, for simplicity we only consider deterministic swapping operations in our setup. 

\begin{figure}[!htb]
    \centering
    \includegraphics[scale = 1.2, clip, trim = {0.1cm, 0, 0.1cm, 0}]{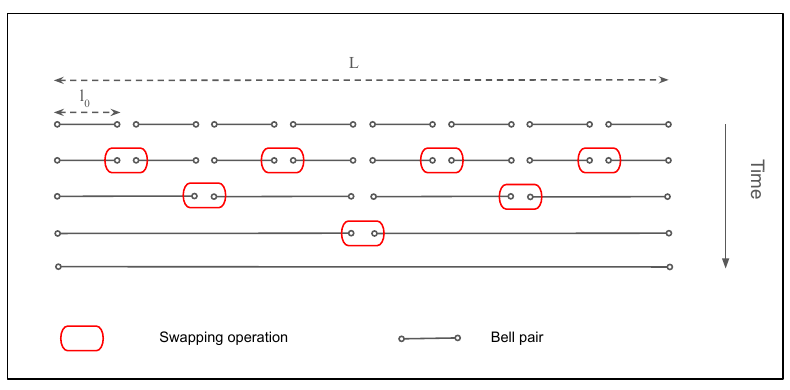}
    \caption{\textbf{Schematic of a nested swapping protocol}. We start with $N$ elementary segments each of length $l_0$. The red boxes represent a swapping operation. Based on the decision variable $\mathcal{D}_i$, distillation may be performed before a swapping operation. Each swapping operation doubles the length of the link. This process is repeated until an end-to-end link is established.}
    \label{fig:nested_swapping}
\end{figure}

In our protocol we use a nested swap strategy based on the Innsbruck Protocol~\cite{dur_quantum_1999, hartmann_role_2007}. The Innsbruck protocol involves a series of entanglement swaps where qubits initially entangled with nearby nodes are used to establish entanglement with more distant nodes through intermediary swaps. This yields a nested structure where the network is divided into $N = 2^n$ segments (see Figure \ref{fig:nested_swapping}). This nested swap procedure allows for the establishment of long-range entanglement connections between nodes that are not directly adjacent. By recursively applying entanglement swap operations, the protocol facilitates the generation of entangled links across the entire network. Prior works have studied swapping schedules other than nested swapping like Swap ASAP~\cite{kozlowski_designing_2020, pouryousef_analysis_2024}, sequential generation and swapping~\cite{xiao_connectionless_2024, pouryousef_analysis_2024}, hybrid strategies~\cite{bacciottini_redip_2024} among others. However, nested schemes perform better than several other swapping schemes especially in settings where the swapping operations are probabilistic, and repeaters are equipped with distillation, by providing an entanglement distribution rate that decays polynomially rather than exponentially in distance~\cite{duan_long-distance_2001}. Moreover, a nested swapping schedule allows for node synchronization for generation, swapping, and distillation operations. It is because of these reasons, coupled with an ease of analysis, that we have chosen a nested swapping schedule for our protocol.

Our protocol diverges from the standard Innsbruck protocol in two ways --- (1) distillation may or may not be performed at each level depending on the expected quality of the Bell states (See subsection~``\nameref{subsec:recursive_calcs}'') (2) all links are created in parallel with multiple links shared between two stations, in a single burst with no interaction between bursts. In our protocol, we consider swaps as deterministic operations which allows us to save on the associated classical communication time costs. However, it is important to note that unless one has perfect elementary links, swaps even in the case of perfect gate and measurement operations, cause an exponential decay in the fidelity with each swap~\cite{briegel_quantum_1998}. In the absence of a means to improve fidelity, this exponential decay makes quantum networks based solely on swaps impractical for long range communications. Subsections ``Gate operations'', and ``Measurement Operations'' in the Supplementary Methods explain the models used for gate and measurement operations, and Supplementary Equation (9) has been used for modeling the swapping operations.

\subsubsection{Distillation}\label{subsec:Distillation}
The degradation in entanglement fidelity due to imperfect operations (such as swaps) or imperfect memories used for storing the entangled qubits can be mitigated by using distillation. In a distillation operation a larger number ($n $) of Bell pairs are transformed to achieve a smaller number ($k$) of higher fidelity Bell pairs. While significant improvements have been made in the field of distillation~\cite{rozpedek_optimizing_2018, krastanov_optimized_2019, jansen_enumerating_2022, goodenough_near-term_2024}, in this paper, for ease of analysis,  we use one of the basic schemes called the DEJMPS protocol ($n = 2, k = 1$)~\cite{deutsch_quantum_1996} (See Supplementary Equation (7) in the ``Distillation'' subsection in Supplementary Methods).  The DEJPMS protocol is typically employed in an iterative fashion until a threshold fidelity is achieved or a threshold number of Bell pairs have been exhausted to create a higher fidelity pair. However, since most of our evaluation focuses on high input fidelity Bell pairs, we have considered only a maximum of a single round of distillation before swapping. Our protocol is easily modified to account for multiple rounds of distillation i.e. performing as many distillation rounds until the fidelity of all available Bell pairs reach above the threshold fidelity, or we run out of multiplexed Bell pairs available.

A critical consideration is the temporal overhead introduced by executing DEJMPS. In deterministic distillation protocols, classical communication is needed to relay Pauli correction information between parties. However, this exchange does not introduce latency, allowing operations to proceed without delay. In contrast, probabilistic distillation protocols impose stricter timing requirements, as the success or failure of the distillation must be communicated to determine subsequent actions. This requirement can create a significant bottleneck in two-way architectures. To address these temporal costs, one approach is to operate in a ``blind'' mode~\cite{hartmann_role_2007}. However, in our protocol, the distillation operations are conducted with ``informed'' decisions~\cite{razavi_quantum_2009, razavi_physical_2009}.

Deterministic distillation protocols might offer better performance than the probabilistic protocols like DEJMPS on metrics such as secret-key rates and memory usage over time. However, deterministic distillation protocols typically require a large number of input entangled states for creating a higher fidelity state.  Another possibility is to utilise a combination of deterministic and probabilistic distillation schemes with high fidelity output states.

Another important consideration in two-way repeaters is the decision whether to distill or not before performing a swap. Given a higher initial fidelity, it is possible to perform multiple swaps and increase the length of the link before the fidelity drops to a level where distillation might be required. For a nested two-way protocol, this decision is made at each nesting level, determining whether one or multiple rounds of distillation are required to deliver Bell pairs end-to-end. Typically, this decision is not reactive, and does not depend on real-time measurement outcomes or classical information from other network nodes. Instead, it is pre-computed prior to run-time based on factors such as the expected fidelity, number of multiplexed channels available, operational noise, and the number of repeaters.

As discussed in subsection~``\nameref{subsec:multiplexing}'', and further explained in subsection~``\nameref{subsec:recursive_calcs}'', it is possible to keep track of the probability distribution of the number of Bell pairs. This probability distribution can be further optimised over the decision to distill based on a key service metric (such as the secret-key rate per shot or fidelity threshold). In the proposed protocol, this decision parameter is pre-determined for all nested levels, and is taken to be a static network-wide agreement.  This decision can be made using any rule that might be suitable to the application, and the metric to be optimised. As examples, we have included two case scenarios in our analysis for deciding when to perform distillation --- (1)  comparing expected secret-key rates with and without distillation for a chain with $2^{n-i}$ segments, where $i$ is the nesting level, and $n = \log_2 N$ for an $N$ elementary segment linear network (See equation~\eqref{eq:SKR_Decision} in subsection~\sectionref{subsec:recursive_calcs}). This is primarily driven by the fact that secret-key rate combines both throughput and fidelity into a single metric making it a useful metric to optimise. This rule has been referred to in this manuscript as the $SKR$ rule. (2) comparing the fidelity of links to a pre-determined fidelity threshold, where the decision to distill is contingent upon the link quality being less than the threshold (referred to in this manuscript as the $F_{th}$ rule). This approach could potentially be useful in a scenario where the quality of links above a certain threshold is desired as a service metric~\cite{kozlowski_designing_2020}. Both of these policies can be further optimised with an objective to maximise the end-to-end secret-key rate or number of Bell pairs while considering various possible distillation schedules and protocols.
Furthermore, since we consider distillation, we need to keep track of the Bell pairs sacrificed when a round of distillation is performed. For a general $n'$-to-$k'$ distillation scheme, each successive distillation round reduces the number of available Bell pairs by a factor of at least $1 / \lceil n'/k' \rceil$ --- see Supplementary equations (14) and (15) in the Supplementary Materials for details.


\subsubsection{Quality of Memory}

The temporal costs associated with classical communication with distillation requires Bell states to be held in long-lived memories that do not undergo significant decoherence. Degradation in memory quality is usually characterised by $T_1$ and $T_2$ times. The $T_1$ time denotes the thermal relaxation time - the time it takes for the excited state $\ket{1}$ to relax back to the ground state $\ket{0}$. The $T_2$ time is the dephasing time that captures the loss of coherence due to dephasing in a quantum memory. In this manuscript, we only consider dephasing noise (See the ``Memory Decoherence'' subsection in the Supplementary Methods in the Supplementary Materials).


\subsubsection{Termination}
The protocol concludes once one or more Bell pairs are successfully established between the end stations. However, there may be cases where repeater stations lack enough Bell pairs to perform distillation. We explore various termination strategies for such scenarios when the number of Bell pairs in a segment drops below a certain threshold ($R_i$) for any nesting level $i$. If the static distillation schedule --- predetermined based on a distillation rule such as a fidelity threshold $F_{th}$ or the $SKR$ rule outlined in this manuscript --- requires at least one round of distillation at the current or higher nesting levels, the protocol must adapt accordingly. Using this framework, we propose three potential termination strategies:

\begin{itemize}
    \itemsep0em
    \item \textbf{Strategy 1:} Repeater stations in the affected segment send classical messages instructing Alice, Bob, and all intermediate repeaters to halt all operations related to the burst. As these messages propagate, the repeater stations release the memory resources associated with the burst.
    \item \textbf{Strategy 2:} A variation of the first strategy involves the repeaters performing entanglement swaps and notifying their counterpart stations to perform additional swaps without distillation. Here, counterparties refer to repeaters that share a Bell pair with the initiating repeater. This approach allows repeaters holding $k$ links to proceed similarly, ultimately establishing $k$ end-to-end links. However, because this strategy bypasses the static distillation schedule, the established links will likely be of lower quality.
    \item \textbf{Strategy 3:} Another option is to allow unaffected segments to proceed without interruption. Specifically, repeater stations can continue performing distillation and swapping operations in segments where the number of Bell pairs $k \geq R_i$. Repeater stations learn about the failure on a segment as and when they do, minimizing classical communication time compared to Strategy 1. Also, this strategy requires fewer resources than Strategy 2, reducing the need for gates and other operations, while freeing up memory for future bursts.
\end{itemize}

The threshold $R_i$ can be set based on the minimum number of Bell pairs required for the chosen $n'$-to-$k'$ distillation scheme (such that $R_i \geq n'$). Alternatively, this threshold may be optimised using the static distillation decision schedule, ensuring that termination does not occur at any nesting level. We have selected \textbf{Strategy 3} because it minimises classical communication time, potentially avoids unnecessary gate and measurement operations, and frees up memory resources for future bursts. However, this strategy sets a lower bound on performance. While computationally challenging, a more efficient termination strategy that optimises resource usage could be developed in future work.


\subsection{Recursive Formulation of the Probability Distribution}\label{subsec:recursive_calcs}

We consider a linear network with $N=2^n$ links. Let $M = m 2^{n + 1}$ denote the number of multiplexed channels available at each elementary link in a single shot, with $m \ge 1$. note, this assumption can be relaxed if distillation is not required on all levels to $M$ taken to be lesser than $m 2^{n + 1}$. We consider a nested pumping distillation protocol that performs at most one distillation round at each level.

Let $Y_i$ denote the number of Bell pairs on a segment at level $i$ with $p_{i,k} = P(Y_i = k)$ denoting the probability of having exactly $k$ Bell pairs at level $i$. Let $\pi_0$ denote the probability that a link-level Bell pair generation attempt succeeds, and let $\overline{\pi}_0 = 1 - \pi_0$. Then,
\begin{align}\label{eq:basic_prob_generation}
    p_{0,k} = \binom{M}{k} \pi_0^k \overline{\pi}_0^{M - k}, \quad k= 0,1,\ldots, M\ .
\end{align}

\subsubsection{Multiplexing with Exactly One Success - 2GNC}\label{subsec:subsec_2gnc}
For 2G-NC, the focus is on creating at least one bell pair in all segments. Further, using a similar setup assumed in ~\cite{muralidharan_optimal_2016}, it is assumed that there are no distillation operations, and since swapping operations are deterministic we perform a network wide swap simulatenously on the single link created, thus creating an end-to-end bell pair with a success probability, 
\begin{align}
\label{eq:2gnc_prob_generation}
    \textit{Success Prob. }_{(2GNC)}  &= \Big(\sum_{k=1}^M p_{0,k}\Big)^N, \quad k= 0,1,\ldots, M\nonumber\\
     &= \Big(\sum_{k=1}^M \binom{M}{k} \pi_0^k \overline{\pi}_0^{M - k}\Big)^N \text{, using equation \eqref{eq:basic_prob_generation}}\nonumber\\
     &= \Big(1 - p_{0,0}\Big)^N \nonumber\\
     &= \Big(1 - \overline{\pi}_0^M\Big)^N
\end{align}

\subsubsection{Multiplexing with more than one success along with distillation}\label{subsec:dejmps_recursive}

In order to calculate $p_{i,k}$, we first determine the effect of a distillation operation at level $i$ whenever it is performed. This is captured by $q_{i,k}$,
\begin{align}
    q_{i,k} = P(X_i = k) = \sum_{j=2k}^{M/2^i} p_{i,j} \binom{\lfloor j/2 \rfloor}{k} d_i^k \overline{d}_i^{\lfloor j/2 \rfloor - k}, \quad i \in \{0,\cdots, n\} ;k = 0,1,\ldots, \lfloor M/2^{i+1} \rfloor.
\end{align}
where $ X_i$ denote the number of Bell pairs produced by one distillation step at level $i$ when performed, and $d_i$ the probability of a successful distillation step. This equation can be extended to reflect the case when no distillation is performed at that level. That is,
\begin{align}
    q_{i,k} = P(X_i = k) = 
    \begin{cases}
            \sum_{j=2k}^{M_i} p_{i,j}\binom{\lfloor j/2 \rfloor}{k} d_i^k \overline{d}_i^{\lfloor j/2 \rfloor - k}, &\mathcal{D}_i = 1 \\ 
            p_{i,k}, & \mathcal{D}_i = 0 
    \end{cases},\quad i=0,\ldots,n ;k  = 0,1, \ldots,\lfloor M_i/2 \rfloor
\end{align} 
where $M_i = \lfloor M / 2^{\sum_{j=0}^{i-1}\mathcal{D}_j} \rfloor$ for $i > 0$, and $\mathcal{D}_i$ is the indicator function for distillation,
\begin{align*}
\mathcal{D}_i &= 
    \begin{cases}
        0,\text{ if no distillation at level } i\\
        1,\text{ if distillation at level } i 
    \end{cases}, \quad i = 0, \dots, n
\end{align*}
To note, for the current analysis we have a static value of $\mathcal{D}_i = 0$, when $i = n$. Now,
\begin{align}
    p_{i,k} = 
    \begin{cases}
        (q_{i-1, k})^2 + 2q_{i-1, k} \sum_{j= k+1}^{M_{i}}q_{i-1,j}, & \mathcal{D}_i = 1 \\
         (p_{i-1, k})^2 + 2p_{i-1, k} \sum_{j= k+1}^{M_{i}}p_{i-1,j}, &\mathcal{D}_i = 0 \\
    \end{cases},\quad i=1,\ldots,n; k = 0,1, \ldots, M_i.
\end{align}We now consider a protocol that terminates whenever $Y_i < R_i$ where $R_i$ denote a termination threshold for each level $i$ segment. 
Let,
\begin{align*}
    r_i &= \text{Probability of reset at level $i$ conditioned on reaching level $i$},\\
    p'_{i,k} &= P(Y_i = k | \text{ no reset at levels } 0,1,\dots,i),\\
    q'_{i,k} &= \text{Probability of having $k$ distilled pairs conditioned on reaching level $i$}.
\end{align*}
Now,
\begin{equation}
        r_0 = \sum_{j=0}^{R_0-1} \binom{M}{j} \pi_0^j \overline{\pi}_0^{M - j}
\end{equation}
and
\begin{align}\label{eq:p_prime_0}
    p'_{0,k} = 
    \begin{cases}
        \big(\binom{M}{k}\pi_0^k \overline{\pi}_0^{M - k}\big)/(1 - r_0), &  k \ge R_0\\
        0, &   k < R_0
    \end{cases}, \quad \forall k \in \{0,1, \dots, M\}. 
\end{align}

Now, as defined earlier,
\begin{align}\label{eq:q_prime_dejmps}
    q'_{i,k} &= P(Y_i = k | \text{ no reset at levels } 0,1,\dots,i),\nonumber\\
    &= \begin{cases}
             \sum_{j = 2k}^{M_i} p'_{i,j} \binom{\lfloor j/2\rfloor}{k}d_i^k \overline{d}_i^{\lfloor j/2 \rfloor - k}, & \mathcal{D}_i = 1 \\
            p'_{i,k}, &\mathcal{D}_i = 0 \\
    \end{cases}, \quad \forall k \in \{0,1,\dots, \lfloor M_i/2 \rfloor.\}
\end{align}

\begin{align}\label{eq:conditional_reset_prob}
    r_i = 
    \begin{cases}
        \sum_{l = 0}^{R_{i - 1}-1}\big((q'_{i-1, l})^2 + 2q'_{i-1, l} \sum_{j = l+1}^{M_{i}} q'_{i-1,j}\big), & \mathcal{D}_i = 1 \\
       0, & \mathcal{D}_i = 0 \\
    \end{cases}, \quad \text{s.t. } R_i \le R_{i-1} \quad \forall i \in \{1, \dots, n\}
\end{align} 

Now,

\begin{align}
\label{eq:p_prime_i}
    p'_{i,k} = 
        \begin{cases}
            \big((q'_{i-1, k})^2 + 2q'_{i-1, k} \sum_{j= k+1}^{M_{i}}q'_{i-1,j} \big)/(1 - r_i) , & \mathcal{D}_i = 1 \quad \& \quad k \ge R_i\\
            (p'_{i-1,k})^2 + 2p'_{i-1,k}\sum_{j= k+1}^{M_{i}}p'_{i-1,j} , & \mathcal{D}_i = 0 \quad \& \quad k \ge R_i\\          
            0 , & \quad \quad k < R_i
        \end{cases}, 
        \quad \forall k \in \{0, \dots, M_i\}
\end{align} 

Now, let $f_i$ be the probability of reset at level $i$, 
\begin{align}\label{eq:reset_probability}
f_i =
    \begin{cases}
        1 - (1 - r_0)^N &,  i = 0 \\
        (1 - (1 -r_i)^{{\frac{N}{2^i}}})\prod_{j = 0}^{i-1} (1 - r_j)^{\frac{N}{2^j}} &, \text{otherwise}\\
    \end{cases}
\end{align}\\
The decision to distill $\mathcal{D}_i$ can be computed using any criterion best suited for the application. As examples, we have considered two conditions - (1) $F_{th} \underset{\mathcal{D}_i = 0}  {\overset{\mathcal{D}_i = 1}{\gtrless}} F_{i}$ where $F_i$ is the fidelity at the $i^{th}$ level and $F_{th}$ is a pre-decided threshold fidelity, (2) $SKR^\uparrow_i  \underset{\mathcal{D}_i = 0}  {\overset{\mathcal{D}_i = 1}{\gtrless}} SKR_i$, where $SKR_i$ is the secret-key rate at level $i$ without distillation, and $SKR^\uparrow_i$ is the secret-key rate at level $i$ after distillation. Using Supplementary equation (12) in the Supplementary Materials, this $SKR$ based decision rule can be further elaborated as, 
\begin{align}\label{eq:SKR_Decision}
     SKR^\uparrow_i & \underset{\mathcal{D}_i = 0}  {\overset{\mathcal{D}_i = 1}{\gtrless}} SKR_i\nonumber\\
      \text{i.e. } \quad\quad r_{secure}(\rho_i^\uparrow)\cdot \mathbb{E}(Y_i^\uparrow) & \underset{\mathcal{D}_i = 0}{\overset{\mathcal{D}_i = 1}{\gtrless}} r_{secure}(\rho_i)\cdot \mathbb{E}(Y_i),
\end{align}
where $\rho_i$ is the two-qubit state shared between two parties before distillation at level $i$, $Y_i^\uparrow$ denote the number of Bell pairs on a segment at level $i$ after one round of distillation, $\rho^\uparrow_i$ is the state after distillation at level $i$, $r_{secure}(\rho)$ denote the secret-key fraction for a two-qubit state $\rho$, and 
\begin{align}\label{eq:expected_bp_endtoend}
    \mathbb{E}(Y_i) &= \prod_{j = 0}^{i} (1 - r_j)^{\frac{N}{2^j}} \cdot \sum_{k=R_i}^{M_i} k \cdot p'_{i,k} \nonumber\\
    \mathbb{E}(Y_i^\uparrow) &= \prod_{j = 0}^{i} (1 - r_j)^{\frac{N}{2^j}} \cdot \sum_{k=R_{i}}^{\lfloor M_i/2 \rfloor} k \cdot q'_{i,k} 
\end{align}
To note, for both cases listed above, we do not distill when equality holds.


\subsection{One-way Repeater Architecture}

First proposed by Munro \textit{et al.} 2013~\cite{munro2012quantum} and 2015~\cite{munro_inside_2015}, one-way repeater architectures uses near-deterministic methods to handle loss and operational errors. One fundamental difference between this scheme and a two-way scheme is its requirement for only forward or one-way classical communication. This need for one-way classical communication can further be eliminated if recovery operations on the errors accumulated in the preceding segment are performed at each repeater, resulting in only forward flow of quantum information. To tackle errors, in a one-way scheme, the quantum state to be transmitted is encoded in a logical qubit (qudit) using several physical qubits (qudits). For our analysis, we only consider a Bell pair of which one qubit is encoded and transmitted, while the other stays with the initiating party. Depending upon the nature of errors, there are various quantum error correction codes that can be used to encode the Bell pair and protect against these specific anticipated errors. For our comparison, we focus on Quantum Parity Codes (QPCs)~\cite{munro2012quantum, muralidharan_ultrafast_2014}. QPCs are generalised Shor Codes, and are capable of supporting Teleportation Based Error Correction~\cite{jiang_quantum_2009}. A general form of an $(n,m)$ QPC encodes the logical qubits as $\ket{0}_L = (\ket{+}_L + \ket{-}_L)/\sqrt{2}$ and  $\ket{1}_L = (\ket{+}_L - \ket{-}_L)/\sqrt{2}$, with 
\begin{align*}
    \ket{+}_L = \frac{1}{2^{n/2}} \big(\ket{0}^{\otimes m} + \ket{1}^{\otimes m }\big)^{\otimes n}\ ;
    \ket{-}_L = \frac{1}{2^{n/2}} \big(\ket{0}^{\otimes m} - \ket{1}^{\otimes m }\big)^{\otimes n}\ .
\end{align*}
QPCs can be used to recover any encoded state under erasure noise as long as the following two conditions are met --- (1) at least one qubit must arrive for each sub-block; (2) at least one sub-block must arrive with no loss. QPCs are loss tolerant~\cite{ralph_loss_2005}, which is particularly useful to counter erasure losses in the optical fibre. They can also be prepared fault-tolerantly~\cite{muralidharan_ultrafast_2014}. This error tolerance makes QPCs well suited for the one-way architecture, which must correct for loss and gate errors in the absence of heralding or two-way feedback. The logical teleportation procedure implemented at each repeater allows the encoded Bell state to be re-encoded and forwarded, while also correcting for errors introduced in the previous segment. This process proceeds recursively along the chain, enabling end-to-end entanglement to be generated without the need for any backward classical signalling.

To achieve this feed-forward functionality, a teleportation-based error correction (TEC) scheme is used. In this scheme each repeater node corrects errors from the preceding segment and prepares a clean, re-encoded logical state to be forwarded downstream. This process allows for pipelined, segment-wise error correction and preserves the fidelity of end-to-end entanglement across a chain of repeaters. The TEC protocol works by transferring the quantum state of an incoming encoded block to a freshly prepared logical qubit, using a combination of entanglement, measurement, and Pauli frame updates. The flow of operations at each repeater follows these four steps:

\begin{enumerate}
\itemsep0em
    \item Photon loss detection: A quantum non-demolition (QND) measurement is first applied to the incoming logical qubit $\ket{\psi}_L$ to detect photon losses. This allows identification of missing physical qubits without disturbing the surviving ones.
    \item Preparation of logical Bell pair: Locally, a fresh encoded Bell state is prepared using a $\ket{+}_L$ and a $\ket{0}_L$ block. One half of this pair will eventually carry the teleported state forward.
    \item Entangling operations and logical Bell measurement: The surviving photons of the incoming state are coupled to the $\ket{0}_L$ block via transversal controlled-NOT (CNOT) gates. A logical Bell measurement is then performed: the incoming block is measured in the logical $X$-basis and the entangled block in the logical $Z$-basis, implemented via individual measurements on the physical qubits.
    \item Correction via Pauli frame: Based on the measurement outcomes, an appropriate Pauli correction is applied --- either physically or virtually --- to the untouched $\ket{+}_L$ block. This restores the logical state and yields a high-fidelity outgoing qubit, ready for the next transmission segment.
\end{enumerate}

\begin{figure}[!ht]
    \centering
    \includegraphics[width=0.97\linewidth, trim={0 0 0 0.4cm},clip]{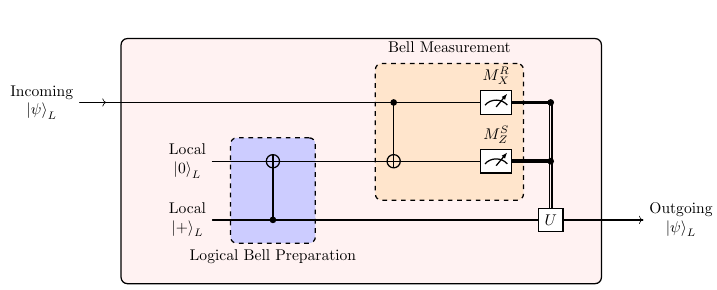}
\caption{\textbf{Teleportation-based error correction (TEC) circuit for the one-way quantum repeater architecture}. $R$ and $S$ denote the incoming and locally prepared code blocks, respectively, each encoded using the same $(n,m)$ quantum parity code (QPC). Syndrome measurements are denoted by $M_{\{X,Z\}}$. The incoming logical qubit $\ket{\psi}_L$, which may have suffered photon loss during transmission through the optical fibre, is coupled to a freshly prepared logical Bell pair via transversal CNOT operations and logical Bell measurements. The local logical states are prepared using a fault-tolerant encoding circuit. Based on the measurement outcomes, a Pauli correction --- denoted by the unitary $U$ --- is applied to the untouched half of the Bell pair, yielding a clean outgoing logical state that preserves the original quantum information. This state is then downloaded onto photonic qubits and transmitted to the next repeater node.}
  \label{fig:oneway_tec}
\end{figure}

A schematic circuit diagram of the TEC operation is shown in Fig.~\ref{fig:oneway_tec}. This segment-wise error correction process repeats recursively across all repeater nodes, ensuring that errors do not accumulate along the chain. For a detailed explanation on syndromes in QPCs, see Namiki \textit{et al.}~\cite{namiki_role_2016}. In our analysis, we assume the codes are prepared fault-tolerantly, and require the same setup as outlined in Muralidharan \textit{et al.} 2014~\cite{muralidharan_ultrafast_2014} and Namiki \textit{et al.}~\cite{namiki_role_2016}. For simplicity, similar to the two-way protocol outlined earlier, we assume spatial multiplexing such that all incoming photons from a block arrive at the repeater at the same time or with negligible time delay. The scheme is compatible with spectral and time-bin multiplexing; however, in those cases, switching losses may need to be explicitly accounted for. Since error correction takes place locally at each repeater, the scheme places more demanding requirements on repeater placements, operation fidelities and speeds, particularly as the code size increases. Nonetheless, for future networks where high-quality operations can be realised, QPC-based one-way schemes offer excellent potential for low-latency transmission. Furthermore, similar to the strategy outlined in~\cite{muralidharan_ultrafast_2014}, we use codes that deliver the highest key rate using the least number of qubits for our comparison. 

\section{Results}\label{sec:Results}
\subsection{Parameter Regime}\label{subsec:parameter_regime}
We develop a model of our protocol in~\nameref{sec:Methods} and use it to compare its performance with that of the one-way scheme. As outlined in subsections~``\nameref{subsec:Distillation}'' and~``\nameref{subsec:recursive_calcs}'' in~\nameref{sec:Methods}, we optimise the decision to perform distillation using a service metric (e.g. secret-key rate or fidelity threshold). Although the model yields entanglement delivery rate and average fidelity, we will use secret-key rate as our metric throughout this section. Our choice of secret-key rate as the primary performance metric is guided by the necessity of establishing a network capable of consistently delivering high-quality Bell pairs at a rapid pace. Secret-key rate combines fidelity (link quality) and entanglement delivery rate (quantity) into a single metric, making it a straightforward choice for measuring performance. See the Secret-key Rate subsection in the Supplementary Methods section of Supplementary Materials for details. 

In Figure~\ref{fig:Performance_withN_loglog}, we show the secret-key rate of the multiplexed two-way protocol across different ranges of coupling coefficients, gate and measurement noise, for distances up to $10^4$ km. We note that the coupling efficiency and gate errors affect secret-key rate in qualitatively different ways. Secret-key rate decreases as coupling efficiency decreases in a uniform manner over all segment lengths whereas increasing gate errors asymmetrically affects more segmented networks.

\begin{figure}[!htb]
    \centering
    \includegraphics[scale = 0.53, clip, trim={0.4cm 0 1.5cm 0}]{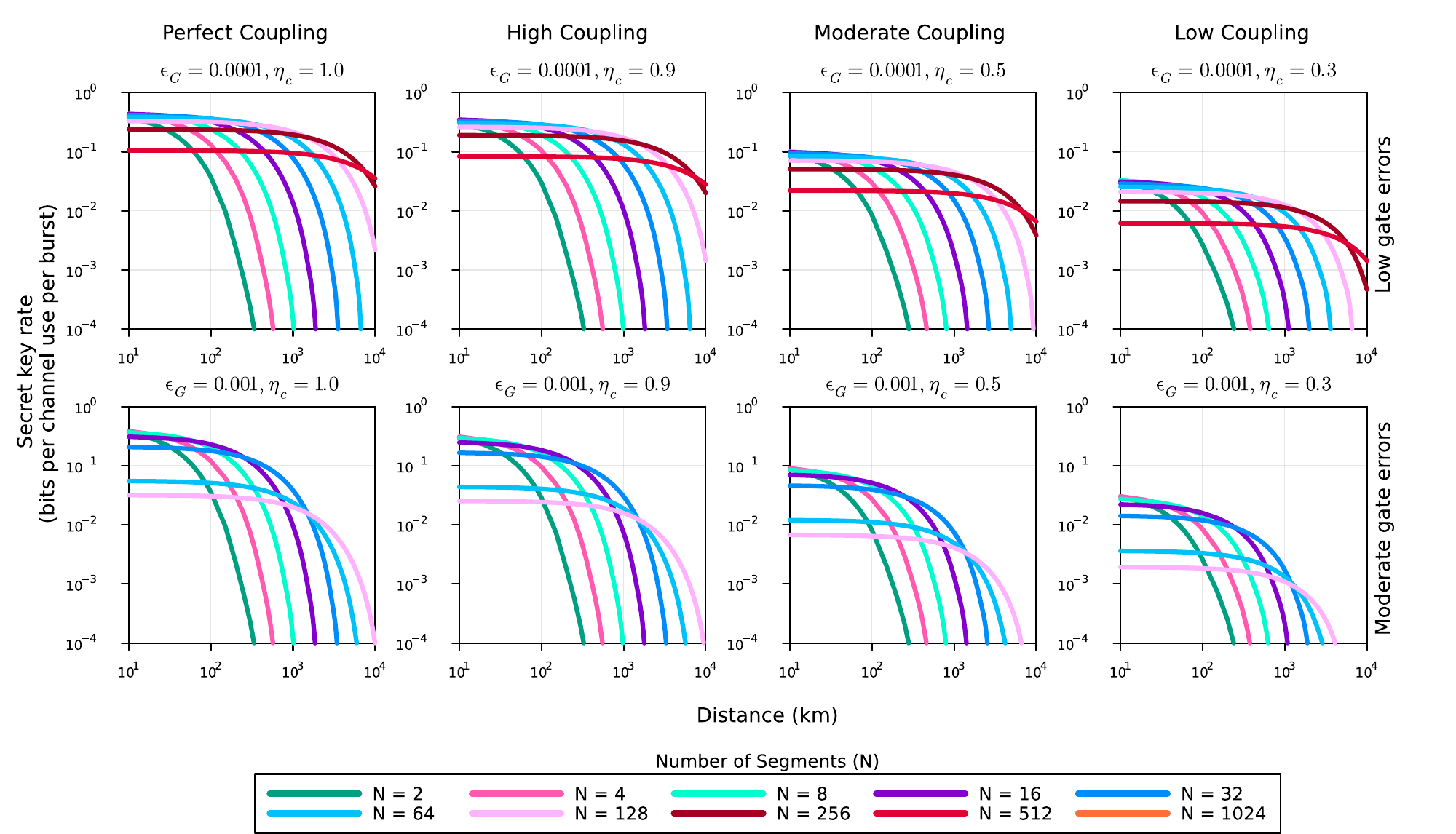}
    \caption{\textbf{Performance of multiplexed two-way protocol (MTP) with distance using secret-key rate as the metric}. The number of segments is shown in different colors and denoted by $N$. The plots in the top row consider a low gate error scenario with a gate error rate ($\epsilon_G$) of $10^{-4}$ or $0.01 \%$, and the bottom row plots show the performance with moderate gate errors ($\epsilon_G$ = $10^{-3}$ or $0.1\%$). The different columns show the performance in different coupling regimes, starting with a perfect coupling ($\eta_c= 1$), with the coupling coefficient reducing when moving from left to right  ($\eta_c \in \{1, 0.9, 0.5, 0.3\}$).  In this setup, we used the protocol based on one-way BB84~\cite{bennett_quantum_2014} Secret-key rate to inform the distillation decision making process, allowing a maximum of one round of distillation at any level of nesting. Also, in this setup, no distillation is performed at the end level of nesting.}
\label{fig:Performance_withN_loglog}
\end{figure}

In this section we outline the parameter regime, and the model assumptions for the results presented in subsections~``\nameref{subsec:performance_eval_SKR}'', and~``\nameref{subsec:costs}'', where we compare the performance of multiplexed two-way protocol (MTP) with one-way schemes. For our comparison, we have selected different Quantum Parity Codes optimised for specific distances for different parameter settings of coupling efficiency, and gate and measurement noise. Furthermore, in our analysis we mainly focus on the parameter regime, demonstrated in prior works to be advantageous for one-way repeater schemes; specifically, as identified by Muralidharan~\textit{et al.}~\cite{muralidharan_optimal_2016}, this regime corresponds to coupling efficiency $\eta_c \ge 0.9$, gate error $\epsilon_G \le 10^{-3}$, and gate time $t_G \le 10^{-9}$s.

For this comparison, keeping in line with the analysis in~\cite{muralidharan_optimal_2016} we assume high initial fidelity (computed as $F = 1 - 1.125\epsilon_G$), gate error $\epsilon_G \in \{10^{-4}, 10^{-3}\}$, measurement error $\xi= 0.25 \epsilon_G$, coupling coefficient  $\eta_c \in \{1, 0.9\}$. Moreover, in this analysis we have chosen a realizable decoherence time with $T_2 = 1 $ seconds. This parameter choice has been experimentally demonstrated in hardware platforms like trapped ions~\cite{pino_demonstration_2021} and Rydberg atoms~\cite{bluvstein2024logical}. In our analysis we consider optimal architectures for both one- and two-way schemes. For the one-way scheme, for each distance, an optimal $(n, m)$ QPC is chosen that minimises the total number of qubits required to deliver unit secret-key, with the search parameters constrained to $n \le 70, m \le 20$ and the inter-repeater spacing constrained between $1$ and $4$ km. For the multiplexed two-way scheme (MTP), a maximum of $1024$ segments, and $1024$ multiplexed channels have been considered, primarily to limit computational costs. For 2G-NC, we use a numerical search for selecting the optimal number of multiplexed channels (ranging between 1 and 1024) that minimised the total number of qubits required over the linear network to deliver unit secret-key. Furthermore, to make this comparison, for both QPC and MTP, only the envelopes of the best performing configuration (in terms of inter-repeater spacing, and additionally for QPC the specific $(n,m)$ code), evaluated using secret-key rate ($SKR$) per channel use per burst as the performance metric, have been considered.


\subsection{Summary of Assumptions}\label{subsec:summary_assumptions}
\begin{itemize}
\itemsep0em
    \item Our protocol assumes unconstrained availability of quantum emitters/memories at all repeaters. In our analysis, we treat memory availability as an upper-bound resource, characterised by the total number of memories per repeater ($2M\nu$), where $M$ denotes the number of multiplexing channels and $\nu$ is the source frequency.  This would reduce memory requirements considerably. Note that we require a memory availability corresponding to the worst-case scenario of all links succeeding and all other probabilistic events succeeding. In a nested distillation scheme, the actual number of memories in practice used per cycle depends on the repeater's position in the chain, with repeaters located at half-way typically storing Bell pairs for longer durations and storage times decreasing with each recursive midpoint. Furthermore, this assumes a worst-case scenario where all elementary links and distillation operators succeed. In practice, one would want to have enough memories to guarantee that all states can be kept with sufficiently high probability. Although we recognise that optimisations in swap scheduling could potentially reduce the memory overhead, our present analysis serves as an upper-bound study, with detailed memory scheduling optimisations deferred to future work.
    \item We assume high cooperativity for the cavity-enhanced memories. Cavities have shown promise in realizations of quantum networks~\cite{reiserer_cavity-based_2015, brekenfeld_quantum_2020}, enabling implementations of efficient multi-qubit gates~\cite{borne_efficient_2020, asaoka_optimization_2021, solak2024universal}, fast storage and readouts~\cite{kollath2024fast}.
    \item We assume switching to be perfect, for both inter-memory connectivity at the repeater, and in case of spectral or temporal multiplexing, the switching required to connect incoming photons to the appropriate memory. Furthermore, we do not account for delays associated with performing CNOT gate operations between any two qubits. These delays costs can be non-trivial with current hardware technology especially if the relative distance between the selected qubits on the register is large. The primary reason for not accounting for these delays is that we assume fast gate operations for both one-way and two-way schemes. Both one-way and two-way schemes will be proportionally hit by temporal costs associated with these two-qubit gate operations. 
    \item We assume detectors are perfect, and that the probability of success of a BSA is exactly $1/2$ at the midpoint stations. 
    \item We assume that detectors and quantum memories can be reset in a time smaller than the inverse of the source frequency $\nu$. We further assume that qubit readout times are sufficiently fast such that they are negligible with respect to the inverse of the source frequency $\nu$. This allows for pipe-lined operations where the only bottleneck is the source's ability to generate bursts. This assumption is motivated by the requirement for the network to operate in a steady state. In our model, a continuous and robust stream of end-to-end Bell pairs is generated by ensuring that memories are quickly reset once a burst of operations has concluded. By requiring the reset time to be smaller than the inverse of the source frequency, we ensure ready availability of memories for the subsequent cycle, thereby reducing unnecessary memory overhead.
    \item We assume that the optical losses in the fibre in transit to be the same for all frequencies in case spectral multiplexing is used. We further assume that the optical fibre does not contribute to any other form of noise except erasure. We assume the speed of light in fibre to be $200,000$ km/s. 
    \item We assume deterministic swap gates. Several proposals that use high cooperativity cavities have shown potential for achieving such gates~\cite{borregaard_long-distance_2015, borne_efficient_2020}. 
    \item Memory decoherence time $T_2$ is assumed to be $1$ second~\cite{noauthor_next_2020, wang_single_2021, dudin_light_2013}.
    \item We assume fast gate and measurement operations (gate and measurement time $t_G \ll 10^{-9}$ seconds).
    \item Measurement errors $\xi$ are assumed to be a quarter of gate errors $\epsilon_G$ i.e., $\xi = 0.25\epsilon_G$~\cite{muralidharan_optimal_2016}. 
    \item Elementary link fidelity is estimated to be $1 - 1.25\epsilon_G$~\cite{muralidharan_optimal_2016} using depolarised states for elementary link Bell pairs.
    \item Fibre attenuation length has been taken as $20$ km.
\end{itemize}


\subsection{Costs}\label{subsec:costs_listed}

Prior works have mostly focused on memory-constrained regimes, and have considered memories as the most significant cost factor~\cite{razavi_physical_2009, razavi_quantum_2009, muralidharan_optimal_2016}. However, promising developments in multiple hardware platforms since~\cite{noauthor_next_2020, wang_single_2021} have weakened these assumptions. It is critical that better cost metrics be considered to evaluate the performance of different repeater architectures. While a detailed cost analysis (accounting for environmental noise, hardware requirements, labour, physical infrastructure, and software development and upkeep) would be the most appropriate approach, we believe the following high-level metrics can still serve as a guideposts for comparing quantum network deployments:

\begin{enumerate}[a)]
\itemsep0em
    \item Cost of repeater installations including acquisition of land and physical infrastructure, maintenance, temperature requirements among others. This is captured in our metric of the number of repeaters required for delivery.
    
    \item Memory costs, including initialisation costs and residence times. Here, residence time refers to the duration for which a memory remains engaged. Since one-way and two-way repeater architectures require vastly different types of quantum memories in the parameter regime of fast gate operations and readout, we capture these costs using the metric of the number of qubits required to deliver a single Bell pair.

    \item Number of 2-qubit gates, and circuit size and usage. In our analysis we have only considered 2-qubit gates as the appropriate measure, since both QPC and two-way nested schemes will require 2-qubit gate operations.

    \item Number of measurement operations. In~\cite{van_meter_path_2013}, the authors identify the number of measurement operations as a potential candidate for evaluating link costs in the context of routing in quantum networks. The authors use simulations to establish the relationship between measurement count and overall network performance, highlighting this metric's potential for assessing resource consumption when determining the optimal path for data transmission. 

    \item Cost of operating a repeater, including energy and ongoing maintenance. Although not explicitly considered in our current analysis, these operational costs are expected to play a significant role in practical deployments, and a detailed evaluation is left for future work.
\end{enumerate}


\subsection{Performance Evaluation using Secret-Key Rates}\label{subsec:performance_eval_SKR}
\begin{figure}[!htb]
    \centering
    \includegraphics[scale = 0.48, clip, trim={0.1cm 0 0.3cm 0}]{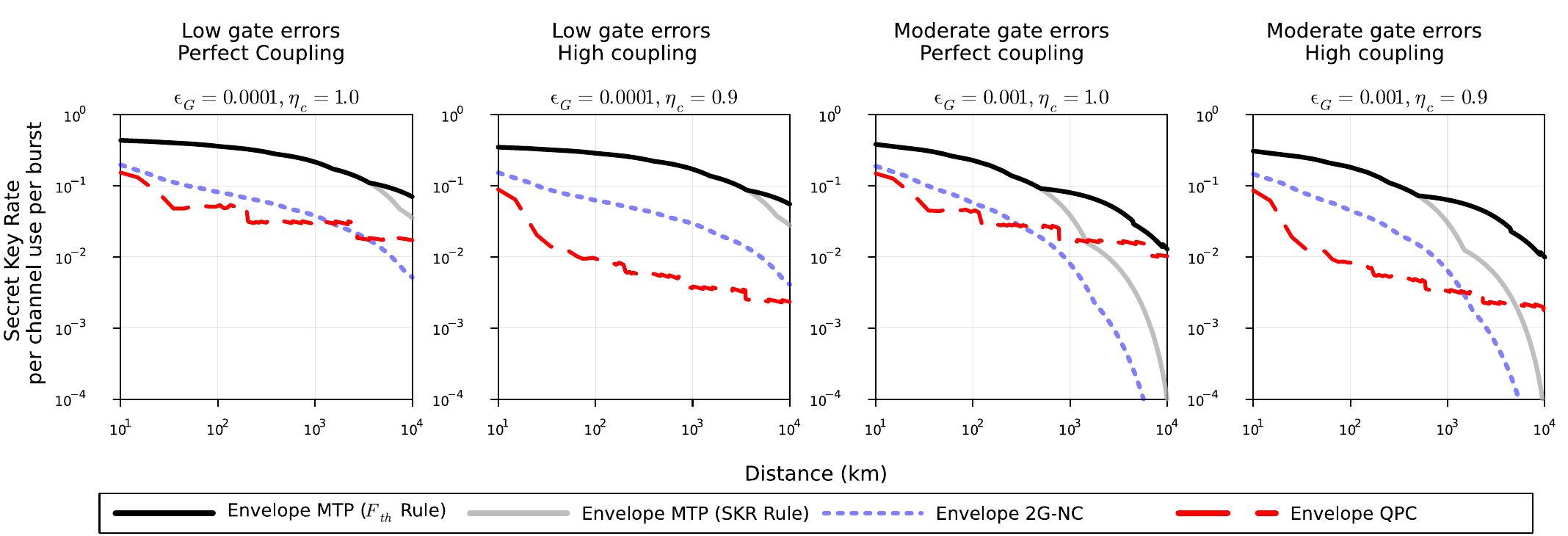}
    \caption{\textbf{Performance comparison between One- and Two-way schemes using the secret-key rate as the metric.} The red dashed line shows the performance by the optimal Quantum Parity Codes (QPC), the blue dotted line represents the optimal performance for the non-encoded second generation scheme `2G-NC', the solid lines are the envelope for the secret-key rates for multiplexed two-way scheme (MTP) with two different distillation rules with black solid line representing the fidelity threshold rule, and the gray solid line representing the $SKR$ rule. To note, for the fidelity threshold rule, we use $F_{th} = 0.95$. For both MTP schemes and the 2G-NC the envelope has been taken over with different number of elementary segments elementary varying between 2 and 1024 . For each distance, a specific $(n,m)$ QPC is chosen optimizing for total number of qubits required with the search parameters constrained to $n \le 70, m \le 20$. For the MTP schemes and the 2G-NC, a maximum of $1024$ multiplexed channels have been considered. Compared to the QPC and the 2G-NC, the MTP schemes deliver better secret-key rates per channel use per burst in all parameter regimes.}
    \label{fig:SKR_comparison}
\end{figure}

We use two different flavours of the multiplexed two-way protocol (MTP) based on two different distillation decision rules --- (1) a rule based on secret-key rate (2) a rule based on fidelity threshold (See~\sectionref{subsec:dejmps_recursive}[sec:Methods] and equation~\eqref{eq:SKR_Decision} for details), where we distill when the fidelity drops below the threshold value. For the fidelity threshold, we have used the threshold of $0.95$, based on a visual search on a small set of threshold values (See Supplementary Note 1 in Supplementary Materials for performance plots using fidelity thresholds other than $0.95$). As shown in Figure~\ref{fig:SKR_comparison}, the MTP repeater schemes outperform the protocol based on the optimal Quantum Parity Codes for all considered parameter regimes. These differences in performance range between one to two orders of magnitude depending on the gate errors and coupling efficiencies considered. In the case of moderate gate errors and long distances, understanding this gain in the context of associated costs, as analysed in~\sectionref{subsec:costs}[sec:Results], is important. We also find that the MTP outperforms the multiplexing protocol aimed at single elementary link generation (referred to as 2G-NC in~\cite{muralidharan_optimal_2016}) across the entire parameter regime considered in the manuscript. To note, both the $SKR$ and the $F_{th} = 0.95$ rule are probably non-optimal, and an optimal distillation schedule can be achieved using a numerical search. Furthermore, since we have limited the number of distillation rounds to a maximum of one per nesting level, potential improvements in the performance could be made if this constraint is relaxed.

\subsection{Comparison of Resource Costs}\label{subsec:costs}

\begin{figure}[!htb]
\includegraphics[width = \textwidth]{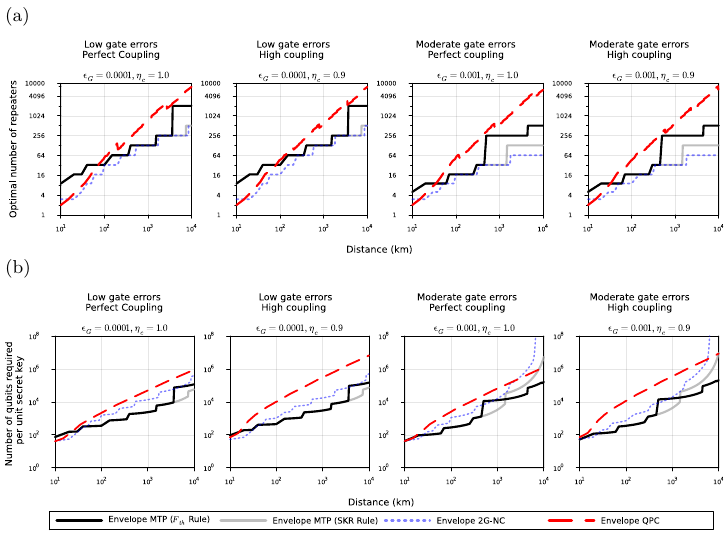}
\caption{\textbf{Comparing hardware costs between one- and two-way quantum repeaters}. Number of (a) repeaters, and (b) qubits required per burst for each unit secret-key delivered for optimal performance for one- and two-way repeater architectures. The red dashed line shows the number of repeaters required for the optimal Quantum Parity Code (QPC), the blue dotted line shows the number of repeaters required for optimal performance for 2G-NC, the black and the gray solid lines are the envelopes for the number of repeaters required for the optimal performing multiplexed two-way schemes (MTP) using a $F_{th} = 0.95$ and a $SKR$ based distillation decision rule respectively. To note, we do not consider the ancilla qubits required for state preparation, or teleportation-based error correction for QPC, and the estimation presented here is a lower bound. For all long distance parameter regimes considered, the MTP requires significantly less number of repeaters, and number of qubits than the QPC. Compared to the 2G-NC, the MTP ($SKR$ rule) scheme require a similar number of repeaters but less number of qubits for delivering unit secret-key. To note, MTP using the $F_{th}$ rule requires slightly more repeaters than the MTP based on the $SKR$ rule, and the 2G-NC protocol, but lower number of repeaters than the QPC.} 
    \label{fig:Repeater_and_Qubit_comparison}
\end{figure}

\begin{figure}[!htb]
\includegraphics[width = \textwidth]{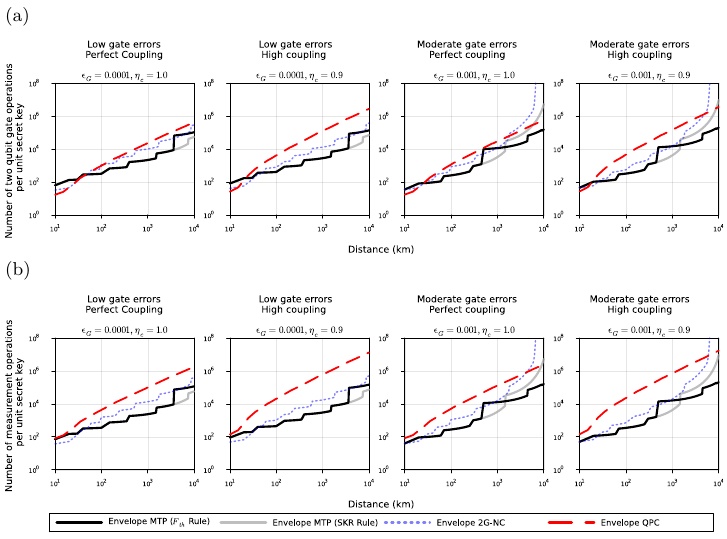}
    \caption{\textbf{Comparing operation costs between one- and two-way quantum repeaters}. Number of (a) two-qubit gates, and (b) measurement operations required per burst for each unit secret-key delivered for one- and two-way repeater architectures. The red dashed line shows the number of repeaters required for the optimal Quantum Parity Code (QPC), the blue dotted line shows the number of repeaters required for optimal performance for 2G-NC, the black and the gray solid lines are the envelopes for the highest deliverable secret-key rate for multiplexed two-way schemes (MTP) using a $F_{th} = 0.95$ and a $SKR$ based distillation decision rule respectively. To note, for QPCs, we have not considered gate operations required for state preparation or gate operations on ancilla qubits, and the estimation presented here is a lower bound. For (almost) all parameter regimes considered, the optimal QPC based protocol require higher number of two-qubit gates and measurement operations than the MTP. }
\label{fig:gate_and_measurement_comparison}
\end{figure}

In this subsection we compare different costs, i.e.~the number of repeaters, number of two-qubit gates, and number of measurement operations. Figures \ref{fig:Repeater_and_Qubit_comparison} and
\ref{fig:gate_and_measurement_comparison} demonstrate that the resources required for the one-way schemes are significantly higher than the equivalently multiplexed two-way schemes. Figure \ref{fig:Repeater_and_Qubit_comparison}(a) compares the number of repeaters required for the optimal secret-key rates shown in Figure \ref{fig:SKR_comparison}. As shown in the plots, the QPC based system requires a significantly larger number of repeaters compared to the multiplexed two-way (MTP) system. This difference in the required number of repeaters becomes more pronounced as imperfections in coupling and gate efficiency increase in the system. Note that 2G-NC requires slightly less or equal number of repeaters compared to the MTP schemes in most parameter regimes. In Figure \ref{fig:Repeater_and_Qubit_comparison}(b), we compare the qubit resources required to deliver a unit secret-key for different gate and coupling efficiencies. In this analysis, we consider a lower bound on the number of qubit resources required for QPC, since we do not consider the ancilla qubits required for state preparation and teleportation-based error-correction. We observe from the graphs that the QPC based system requires more qubit resources for all parameter regimes considered. However, if ancilla qubits are included, it is likely that the two-way scheme will perform even better. It should also be noted that the number of qubits required per unit secret-key delivered has been used as the metric of comparison in Muralidharan \textit{et al.} 2016~\cite{muralidharan_optimal_2016}. Using the number of qubits required as the sole cost metric, it might be straightforward to see the attractiveness of the multiplexed two-way protocol compared to the QPC and the 2G-NC protocol. 
Figures \ref{fig:gate_and_measurement_comparison}(a) and (b) present the estimated number of measurement and 2-qubit gate operations per unit secret-key delivered required to maximise secret-key rate as a function of distance. As with the number of repeaters and qubits, we find that the QPC based one-way scheme requires significantly more gate and measurement operations across most parameter regimes considered with possible exception of low gate errors for distances $\lesssim$ 50 km. For all other considered parameter regimes, these differences in gates and measurement costs range between one and two orders of magnitudes with the MTP posing lower resource requirements compared to both QPC and the 2G-NC.

%

\section{Conclusion}\label{sec:Conclusion}

The rapid development of quantum technologies has spurred efforts to establish robust quantum networks. The choice of repeater architecture significantly impacts the scalability and reliability of these networks. A comprehensive comparison between different repeater architectures is essential to understand their strengths and weaknesses under varying conditions, including error rates, resource availability, and communication latency. Such an analysis can guide the design of practical quantum networks by highlighting where specific architectures excel or falter, and providing insights into the trade-offs between performance and technological complexity.

Pioneering work by Muralidharan \textit{et al.}~\cite{muralidharan_optimal_2016} compared one-way and two-way schemes, identifying parameter regimes where each scheme could be advantageous. However, Muralidharan’s setup assumes a memory-constrained regime and does not utilise the full power of multiplexing. Studies considering multiplexing have focused on maximizing the success probability of a single elementary link or have not incorporated nested purification~\cite{guha_rate-loss_2015, dhara_subexponential_2021, dhara_multiplexed_2022, muralidharan_optimal_2016}. Where such considerations have been made, distillation operations have often been assumed deterministic~\cite{razavi_quantum_2009, razavi_physical_2009}. This study aims to clarify the performance expectations of multiplexed two-way and one-way repeater architectures, providing a framework to make informed decisions when selecting the optimal architecture based on application requirements.

In this manuscript, we consider a two-way protocol that leverages the power of multiplexing with an application-aware decision parameter for distillation. Additionally, we present a thorough evaluation of performance differences between one-way and multiplexed two-way protocols using relevant metrics such as the secret-key rate, number of repeaters, qubits, and gate and measurement operations. Focusing on the regime identified in prior work as favourable to one-way schemes, we demonstrate that the multiplexed two-way repeater scheme, in an unconstrained memory regime,  outperforms one-way schemes even under conditions previously believed to favour the latter. Furthermore, these performance gains can be realised with  lower resource requirement, making two-way schemes a more attractive alternative.

While our findings suggest that multiplexed two-way schemes are potentially a near universal choice across various parameter regimes, the performance achieved in our analysis may be sub-optimal. Our study focused on basic protocols and requires further exploration to identify additional areas for improvement. For instance, our current analysis does not utilise any adaptive mechanisms at the link level for decision-making to optimise overall performance. Additionally, we only considered a basic DEJMPS protocol with a maximum of a single round performed at any nested level. These simplifying assumptions preserve analytical clarity but likely leave room for significant performance improvements. More advanced distillation schemes~\cite{goodenough_near-term_2024, gu2025constant, la2025bayesian} could improve performance while simultaneously lowering resource requirements. Moreover, our distillation scheduling may not be optimal and could be improved to enhance performance and reduce costs. We also assume perfect synchronisation of emissions and neglect delays due to local gate connectivity, which may require further attention in experimental implementations.

Our choice of physical parameters --- such as long memory coherence times, efficient photon coupling, and high-fidelity gates --- reflects recent experimental achievements in platforms like diamond NV centres, silicon T-centres, and other solid-state systems~\cite{bradley2022robust, davidson2023single, zhang2024realization}. Although engineering all these capabilities simultaneously in a scalable system remains challenging, they are not mutually incompatible. Therefore, our parameter set represents an optimistic but feasible near-term scenario that has been partially realised in current experimental platforms.

Building on the framework proposed in this study, future work can focus on improving network performance by introducing optimisations in distillation scheduling or selecting more efficient distillation protocols. Extending this framework to an asynchronous setup or a connection-less protocol might provide interesting insights and potential improvements. Another important direction is to consider architectures involving free-space or satellite-based links, which may differ in coupling efficiencies, loss characteristics, and synchronisation challenges. A detailed exploration of hybrid terrestrial–space quantum networks would help assess the practicality of repeater architectures in global-scale scenarios and is a compelling direction for future study. We recommend exploring these extensions in future research.

\section{Data Availability}\label{sec:data_avail}
All the data presented in this paper can be generated with a customized code~(see Code Availability statement).

\section{Code Availability}\label{sec:code_avail}
The code used to generate this data is available from the corresponding author upon reasonable request.

\section{Author Contributions}\label{sec:author_contributions}
DT formulated the general problem. PM designed the protocols, and conducted evaluations. PM and DT contributed to analysis. KG contributed to discussions. PM wrote the manuscript with inputs from all co-authors. 

\section{Competing Interests}\label{sec:competing_interests}
The authors declare no competing interests.

\section{Acknowledgment}\label{sec:acknowledgment}
We thank Saikat Guha and Alireza Shabani for their valuable comments on the project idea and intermediate results. We are especially grateful to Liang Jiang and Filip Rozp\k{e}dek for their insightful feedback and suggestions on the manuscript. Additionally, we thank the Manning College of Information and Computer Sciences at the University of Massachusetts Amherst for providing access to their High Performance Computation Cluster. The authors acknowledge funding support from the NSF- ERC Center for Quantum Networks grant EEC-1941583, and DOE Grant AK0000000018297.

\section{Supplementary Materials}
Supplementary Materials are available at this \href{https://github.com/mantri-prateek/multiplexed-twoway-protocol}{online repository}.

\afterpage{\clearpage}
\printbibliography

\end{document}